\documentclass[manuscript]{aastex}

\usepackage{fixltx2e}
\usepackage{wasysym}
\usepackage{url}
\usepackage{graphicx}
\usepackage{pdflscape}
\usepackage{afterpage}
\usepackage{epstopdf}

\shorttitle{Analysis of Co-spatial UV-optical STIS Spectra of Planetary Nebula NGC 3242}
\shortauthors{Miller et al.}

\begin{document}

\title{Analysis of Co-spatial UV-optical HST STIS Spectra \\
     of Planetary Nebula NGC 3242\altaffilmark{1}}


\author{Timothy R. Miller\altaffilmark{2}, Richard B. C. Henry\altaffilmark{2}, Bruce Balick\altaffilmark{3}, Karen B. Kwitter\altaffilmark{4}, Reginald J. Dufour\altaffilmark{5}, Richard A. Shaw\altaffilmark{6}, and Romano L. M. Corradi\altaffilmark{7,8}}


\altaffiltext{1}{Based on observations with the NASA/ESA \textit{Hubble Space Telescope} obtained at the Space Telescope Science Institute, which is operated by the Association of Universities for Research in Astronomy, Incorporated, under NASA contract NAS5-26555.}
\altaffiltext{2}{Department of Physics and Astronomy, University of Oklahoma, Norman, OK 73019}
\altaffiltext{3}{Department of Astronomy, University of Washington, Seattle, WA 98195}
\altaffiltext{4}{Department of Astronomy, Williams College, Williamstown, MA 01267}
\altaffiltext{5}{Department of Physics and Astronomy, Rice University, Houston, TX 77251}
\altaffiltext{6}{National Optical Astronomy Observatory, Tucson, AZ 85719}
\altaffiltext{7}{Instituto de Astrof\'{i}sica de Canarias, E-38200 La Laguna, Tenerife, Spain}
\altaffiltext{8}{Departamento de Astrof\'{i}sica, Universidad de La Laguna, E-38206 La Laguna, Tenerife, Spain}


\begin{abstract}
This project sought to consider two important aspects of the planetary nebula NGC 3242 using new long-slit HST/STIS spectra. First, we investigated whether this object is chemically homogeneous by dividing the slit into different regions spatially and calculating the abundances of each region. The major result is that the elements of He, C, O, and Ne are chemically homogeneous within uncertainties across the regions probed, implying that the stellar outflow was well-mixed. Second, we constrained the stellar properties using photoionization models computed by CLOUDY and tested the effects of three different density profiles on these parameters. The three profiles tested were a constant density profile, a Gaussian density profile, and a Gaussian with a power law density profile. The temperature and luminosity were not affected significantly by the choice of density structure. The values for the stellar temperature and luminosity from our best fit model are 89.7$^{+7.3}_{-4.7}$~kK and log(L/$L_{\astrosun}$)=3.36$^{+0.28}_{-0.22}$, respectively. Comparing to evolutionary models on an HR diagram, this corresponds to an initial and final mass of 0.95$^{+0.35}_{-0.09} M_{\astrosun}$ and 0.56$^{+0.01}_{-0.01} M_{\astrosun}$, respectively.
\end{abstract}





\section{Introduction}

Understanding the chemical distribution within the ejected matter that creates a planetary nebula is important for determining a stellar evolution model that accurately describes the progenitor star. Knowing this evolution allows for the proper calculation of stellar yields of elements ejected into the interstellar medium. The ideal candidate to use in a search for spatial variations in the chemical composition is one that has a high surface brightness and is easily resolvable. NGC 3242 is one such planetary nebula. It is bright and extended on the sky, which allows for a detailed comparison at different locations in the nebula of many observed line strengths with their model-predicted values. It also exhibits a multi-shell structure. The brightest features include an inner 28\ensuremath{''}x20\ensuremath{''} shell surrounded by a 46\ensuremath{''}x40\ensuremath{''} halo \citep{rui11}. Thus, the possibility exists that the chemical composition varies from the shell to the halo or from one side of the planetary nebula to the other. This can arise if the halo and shell resulted from mass-loss events occurring at significantly different times during the evolution of the progenitor star or the outflows themselves were inhomogeneous. For stars around 1 solar mass, carbon will be enriched in their nebulae and so is a prime element to look at in NGC 3242. The main goal of this paper (third in the series) is to use spatially resolved HST/STIS spectra presented in Dufour et. al. (2015, hereafter Paper I) to search for positional variations in the abundances of carbon, oxygen, neon, and helium relative to hydrogen and one another.
\\\\Numerous authors have studied the chemical composition of NGC 3242 over the last two decades. \citet{kra06} made long-slit observations in the optical and calculated the abundances for He, O, N, Ne, S, and Cl. The long-slit spectra of \citet{mil02} and the fiber-fed observations of \citet{mon13} covered both optical and near infrared wavelengths. \citet{mil02} calculated abundances for the above six elements plus argon while \citet{mon13} reported all of the elements of \citet{mil02} except neon. Both \citet{hen00} and \citet{tsa03} combined archived IUE ultraviolet data with ground based optical to near infrared long-slit data. The ultraviolet data enabled them to add carbon to the list of elemental abundances for NGC 3242. 
\\\\Out of all of these authors, only \citet{mon13} addressed the idea of chemical inhomogeneity in NGC 3242. They observed the nebula between 3900-7000~\AA~using 6400 fibers onboard the instrument VIMOS-IFU, covering an area of 54\ensuremath{''}x54\ensuremath{''}. They measured accurate oxygen and helium abundances from the most abundant ionic species in this range. For other elements in the study, i.e., nitrogen, sulfur and chlorine, only less abundant ionic species were observable, resulting in large (as high as 6x) discrepancies in the total abundances measured at different locations in the nebula. \citet{mon13} concluded that helium and oxygen were homogeneous throughout NGC 3242. Carbon was not part of that study, unlike in the current paper, since collisionally excited lines of carbon appear only in the ultraviolet. 
\\\\ In addition to the abundance characteristics of NGC 3242, its central star has been studied using many different techniques by numerous authors, e.g.  \citet{fre08}, \citet{pau04}, and Henry et. al (2015, hereafter Paper II), resulting in a broad range of derived effective temperatures (60-95 kK) and luminosities [log(L/L\textsubscript{\astrosun})=2.86-4.01]. Therefore, the second goal of this paper is to better constrain the stellar properties of NGC 3242 through the use of photoionization modeling with the nebular emission lines as constraints.
\\\\We discuss the extraction of the spectra and observational results in \S 2. Section 3 contains the modeling and model results followed by the discussion of these in \S 4. We finish with a summary and conclusions in \S 5.

\section{Spectral Extraction and Observational Results}

\begin{figure}
\caption{The locations of the extracted spectra for NGC 3242 where all seven gratings spatially overlap from STIS. The overall dimensions from 1-9 are 19.8\ensuremath{''}x0.2\ensuremath{''}. In addition to the nine smaller regions, each 2.2\ensuremath{''}x0.2\ensuremath{''}, that were extracted, a Full region spanning all 9 smaller regions was extracted.\newline}
\includegraphics[scale=0.75]{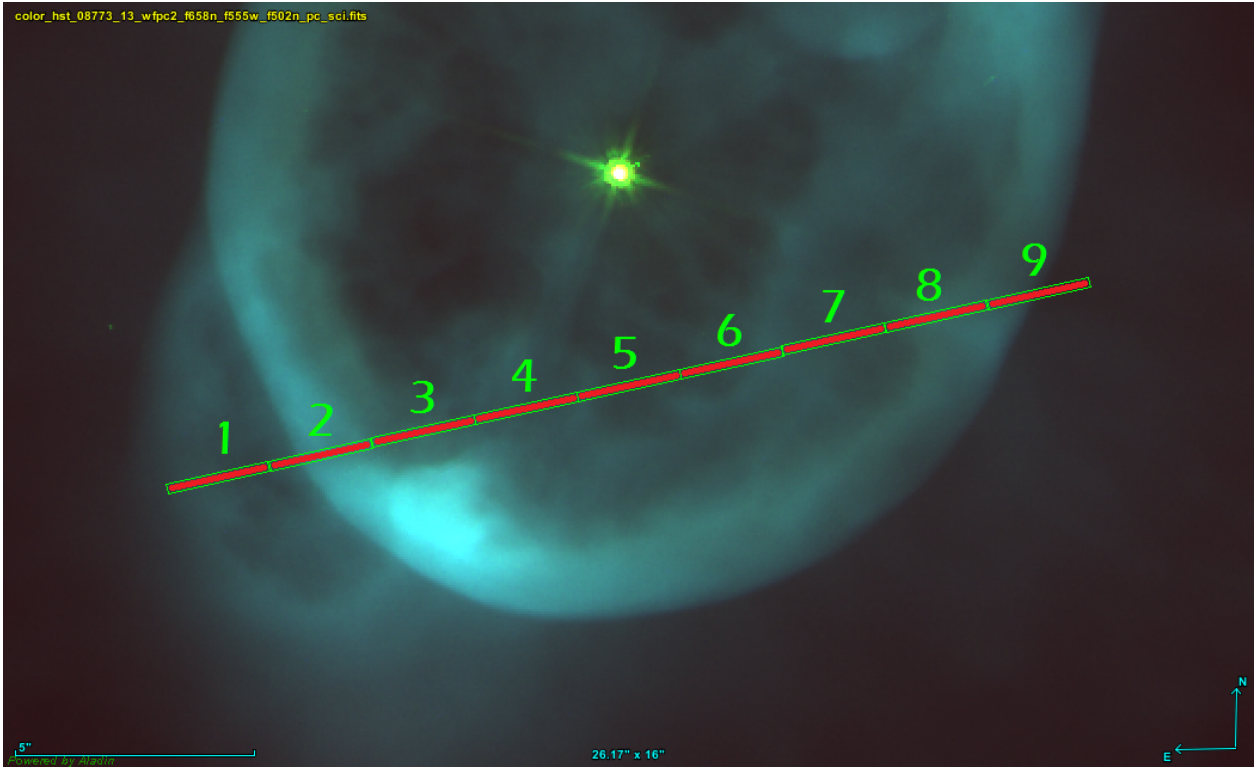}
\label{fig:slits}
\end{figure}

\begin{figure}
\caption{The abundances for carbon, oxygen, neon, and helium for each region as listed in Table~\ref{tab:abun1}. Comparison points HKB2000 and MHK2002 are from \citet{hen00} and \citet{mil02}, respectively. All regions are consistent within errors.\newline}
\includegraphics[scale=0.75]{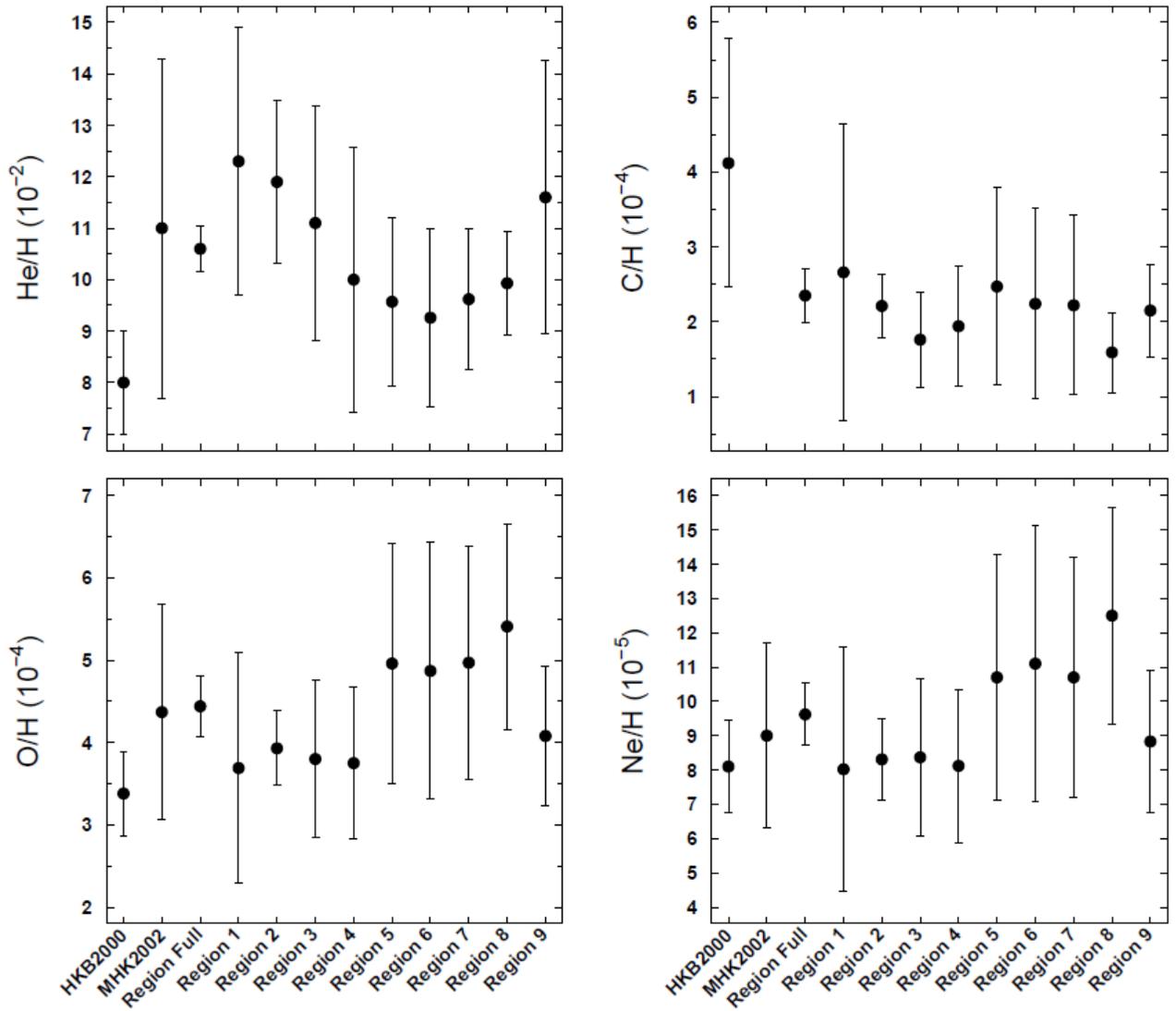}
\label{fig:abun}
\end{figure}

\begin{figure}
\caption{\textit{Left}. The electron temperature for each region based on [O III] as listed in Tables~\ref{ionabun1} \&~\ref{ionabun2}. They are consistent with one another within errors. \textit{Right}. The electron density from C III] for each region along the line of sight as listed in Tables~\ref{ionabun1} \&~\ref{ionabun2}. Regions 3-6 and 8 are consistent within errors with Region Full. Comparison points HKB2000 and MHK2002 are from \citet{hen00} and \citet{mil02}, respectively.\newline}
\includegraphics[scale=0.75]{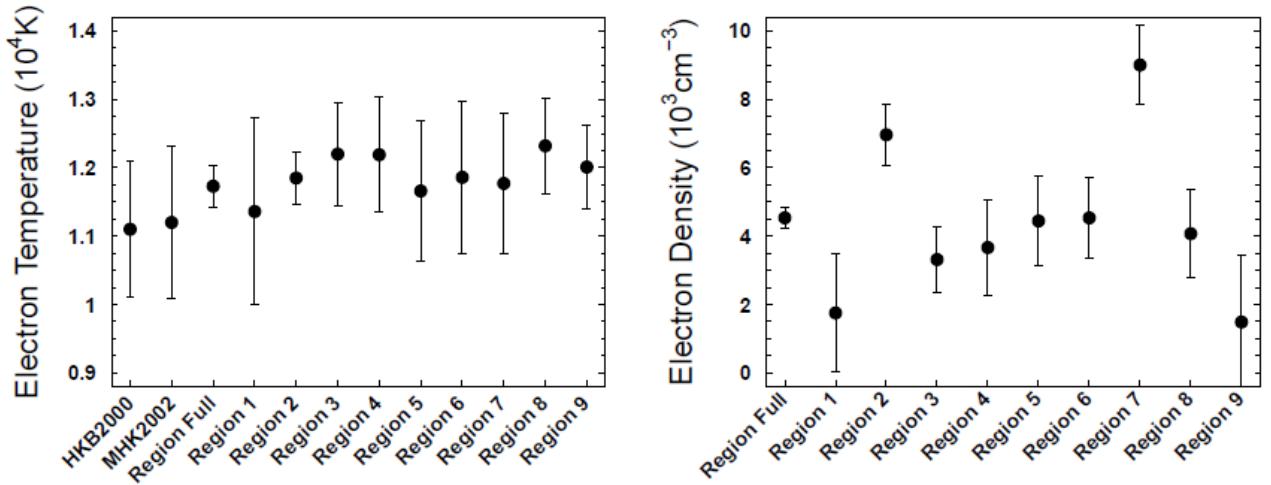}
\label{fig:otempcden}
\end{figure}

A complete description of the observations pertaining to this project can be found in Paper I. To summarize, the spectra used here are part of the observations from the GO12600 spectra cycle 19 program, which is unique in that it consists of co-spatial HST/STIS spectra covering the entire UV-optical-nearIR range from 1150-10,270~\AA~at 0.05$''$ resolution. Extraction of spectra from this data was accomplished by running an in-home Python script which followed the prescriptions in the STIS Data Handbook. The versatility of the script allowed for the division of long-slit observations into smaller regions along the spatial direction. Thus, taking the signal-to-noise into account to maximize the number of emission lines measured, a total of nine regions were chosen, each 2.2\ensuremath{''}x0.2\ensuremath{''}. Our analysis also included consideration of the full region spanning the nine smaller regions. The full region is shown in Figure~\ref{fig:slits} as a green rectangle outlining the nine smaller regions that appear as red filled, green rectangles. Almost all the fluxes of the emission lines were measured in each region using the IRAF\footnotemark \phantom{ }task \textit{splot} by fitting Gaussian profiles. Those that were not, specifically the carbon lines $\lambda$1909 and $\lambda$1907, were measured by summing the observed flux since those lines originate from an M-grating and exhibit flat-topped emission features. As a check for an over-estimation of any one line strength, the summed fluxes from the nine regions had to agree with that of the full region in order for the measurements to be included for further analysis. The uncertainty estimate for each line was calculated from the measured continuum rms noise nearby and the line's FWHM.
\footnotetext[1]{IRAF is distributed by the National Optical Astronomy Observatory, which is operated by the Association of Universities for Research in Astronomy (AURA) under cooperative agreement with the National Science Foundation.} 
\\\\Tables~\ref{emisstab1}~\&~\ref{emisstab2} show our measured and dereddened line intensities by region. The first two columns show the wavelength in \AA~of the emission line and line identification by ion, respectively. The value of the reddening function at each wavelength, f($\lambda$), is shown in the third column. Each pair of columns labeled F($\lambda$) and I($\lambda$) that follows lists the observed and dereddened (with uncertainty) line strengths, respectively, where all values are normalized to F(H$\beta$) or I(H$\beta$) = 100. At the end of each F($\lambda$) column we list the logarithmic reddening parameter, \textit{c}, the theoretical ratio of F(H$\alpha$/H$\beta$), and the observed flux of H$\beta$ in ergs$\cdot$cm$^{-2}\cdot$s$^{-1}$. The values for  F(H$\alpha$/H$\beta$) were calculated using the relevant nebular temperature and density in an iterative loop. The values of \textit{c} are very consistent (rms = 0.01) among the regions and the small differences that are present have negligible effects on the final line strengths and abundances. This argues against the presence of spatially-dependent internal reddening.
\\\\Nebular temperatures, densities, and abundances with errors were calculated from the emission line measurements using the program Emission Line Spectrum Analyzer (ELSA; Johnson et al. 2006). ELSA corrected for interstellar extinction using the function prescribed by \citet{sav79} for optical wavelengths and by \citet{sea79} for the ultraviolet wavelengths. Corrections were also made for the contamination caused by He\textsuperscript{++} recombination lines to the first four hydrogen Balmer lines. Ionic abundance calculations were carried out using a 5-level atom scheme. The propagated uncertainties took into account the input line strength uncertainties as well as the resulting uncertainties in temperature, density and logarithmic extinction. Finally, ionization correction factors (ICFs) were calculated by ELSA using the prescriptions outlined in \citet{kwi01} and Paper I. These ICFs were then applied to the sum of the observed ionic abundances of all elements except carbon to produce elemental abundances. For carbon, the total abundance was taken as the direct sum of the ionic abundances since it was shown in Paper I to be more accurate.
\\\\The ionic abundances and ICF for each observed element are listed in Tables~\ref{ionabun1}~\&~\ref{ionabun2}. The first column lists the ionic species (and wavelength of emission lines in \AA) that relate to values in succeeding columns. The remaining columns contain the ionic abundances for the individual regions. The ICF for each element is listed by region below the ionic abundances. Finally, the two bottom rows of Tables~\ref{ionabun1}~\&~\ref{ionabun2} provide inferred values of [O III] temperature and C III] density. The necessary lines for determining the [S II] density and [N II] temperature were too weak to enable reliable values to be calculated.
\\\\Table~\ref{tab:abun1} contains the elemental abundances of helium, carbon, oxygen, and neon for each region. The first column in Table~\ref{tab:abun1} lists the abundance relative to hydrogen plus the ratios of C/O and Ne/O.  The remaining columns contain the abundance ratios for each region. For reference, the solar values are shown in the last column \citep{asp09}. A comparison by region of each element is shown in Figure~\ref{fig:abun}. Comparison points from \citet{hen00} and \citet{mil02} are also plotted. As can be seen, all of the regional abundances in each panel agree within the errors. Comparing the abundances of our full region to those in \citet{mil02}, all abundances are in agreement, while the helium and oxygen abundances are higher than the values of \citet{hen00}. 
\\\\Similarly, Figure~\ref{fig:otempcden} shows the comparison of each region's [O III] temperature and C III] density. Values from \citet{hen00} and \citet{mil02} are again shown for comparison. As can be seen in the left panel and at the ends of Tables~\ref{ionabun1}~\&~\ref{ionabun2}, the [O III] electron temperatures of the regions range from 11,400-12,300~K but are consistent within errors with the 11,700 K value of the full region. Also, the right panel of Figure~\ref{fig:otempcden} and the ends of Tables~\ref{ionabun1}~\&~\ref{ionabun2} show that along each respective line of sight, regions 2 and 7 have the largest electron densities at 7000 cm\textsuperscript{-3} and 9000 cm\textsuperscript{-3}, respectively, while regions 3-6 and 8 are within the errors of the full region's value of 4500 cm\textsuperscript{-3}. Regions 1 and 9 show the smallest densities of 1800cm\textsuperscript{-3} and 1500cm\textsuperscript{-3}, respectively. The density observed in region 7 appears to be unusually high compared with the much brighter region 2, assuming that the brightness is proportional to the density squared. A plausible explanation is that there is a small knot of enhanced density that is emitting the observed light, though a literature search turns up no corroboration.  

\section{Modeling and Model Results}

\begin{figure}
\caption{The three different density profiles used are illustrated here. The solid red line is the constant density profile. The blue dash line is the Gaussian density profile. The combination of the blue dash line and green dot-dash line is the Gaussian with a power law density profile.}
\includegraphics{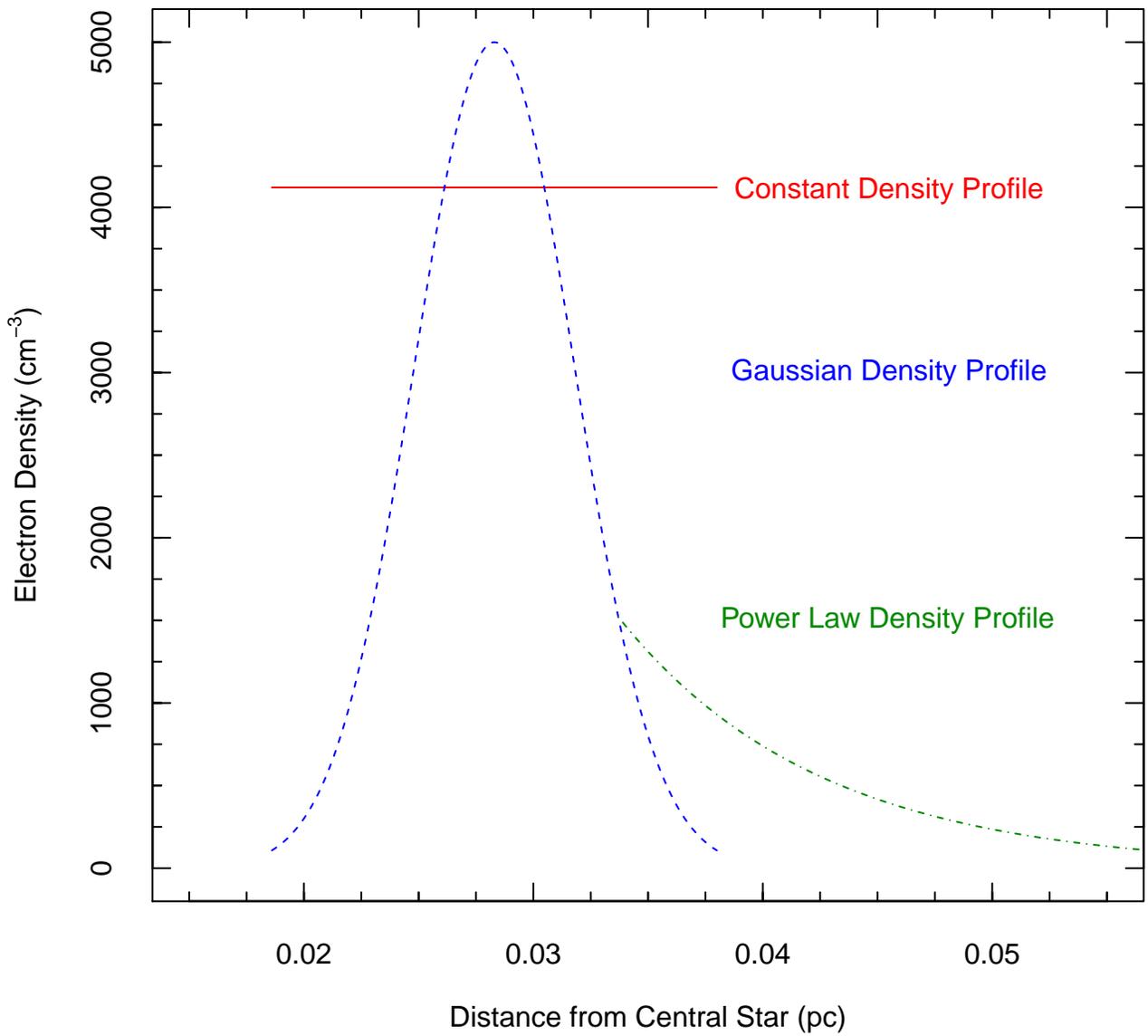}
\label{denshape}
\end{figure}

Photoionization models of NGC 3242 were computed in order to constrain the central star temperature and luminosity using CLOUDY version 13.03 \citep{fer13}. During the computational process, CLOUDY steps outward from the center of the nebula, solving the energy balance and ionization balance equations simultaneously at every point. Three iterations of this process were performed to ensure a steady state solution for the resulting model. Each model employed a Rauch H-Ni stellar atmosphere simulation for the central star spectral energy distribution \citep{rau03}. The Rauch models include line blanketing of all elements on the periodic table from hydrogen to nickel. We assumed a static geometry, spherical symmetry, and no shock heating of the gas. Additional assumptions included a distance to NGC 3242 of 1~kpc and an initial central star temperature and luminosity of 89,000~K and 3450~L\textsubscript{\astrosun}, respectively (Frew 2008, Frew et. al 2016). Finally, the inferred nebular density and abundances from \S 2 were used as initial input into the photoionization models of each region.
\\\\Model line-of-sight line strengths for the emission lines in Tables~\ref{emisstab1}~\&~\ref{emisstab2} were calculated using an in-home C++ program called PlAnetary Nebula Intensity Calculator (PANIC). PANIC used model-generated radial emissivity values paired with their radial distances from the central star to compute each emission line strength. Specifically, PANIC calculated the volume of gas intersected by each region for each radial distance from the model, multiplied it by the appropriate emissivity, and then added up each contribution to determine the total emission for every line. This line emission was then multiplied by the filling factor (the ratio of volume filled by gas to total volume) to account for the clumpiness of the gas in the model. Of the modeled emission lines, at least seven were used to compare with observed line intensities, typically the strongest line for each ion. This ensures equal weighting among ions. Furthermore, values for up to five diagnostics\footnotemark, which are particularly sensitive to nebular properties or the central star's temperature and luminosity, were calculated in order to help break the degeneracy among models whose predicted line strengths otherwise closely matched the observed line strengths.  
\footnotetext[2]{([O III] $\lambda$5007+$\lambda$4959+$\lambda$3727)/H$\beta$, ([O II] $\lambda$3727)/([O III] $\lambda$5007), (He II $\lambda$4686)/(He I $\lambda$5876), [O III] ($\lambda$4363/$\lambda$5007), and C III] ($\lambda$1909/$\lambda$1907)}
\\\\The agreement between model-generated and observed line strengths and diagnostic values was evaluated by calculating a total rms via the expression $\sqrt{\frac{1}{N}\sum_{1}^{N}{(1-\frac{model}{observed})^2}}$, where N is the total number of lines and diagnostics (15 for Region Full). The diagnostics included also act as a weighting system for the more important lines. Similarly, the total rms for observed line strengths and diagnostics was calculated by substituting the uncertainty in place of the model value. This observed rms was used to assess error estimates on input parameters described later. The model with the lowest rms was considered the best model. 
\\\\To test the validity of the described rms method for assessing the best model, we also used the method of~\citet{mor09}. In their method, a quality factor, \textit{Q}, is calculated by the expression $\frac{log(\frac{model}{observed})}{log(1+RelativeError)}$ where RelativeError is $\frac{uncertainty}{observed}$. Minimizing this \textit{Q} value instead of the rms resulted in a small difference in the best fit stellar temperature and luminosity of only 200 K and 0.011 dex, respectively.
\\\\Three different density profiles were tested during our analysis: a constant profile, a Gaussian profile, and a Gaussian with a power law profile. The constant density profile has a defined inner and outer radius and a single density throughout the gas. However, it is likely to be the least appropriate profile of the three, given the multi-shell structure of NGC 3242 and the rigid boundary conditions. The Gaussian profile, on the other hand, relaxes the sharp boundary cutoff and constant density condition. Yet, a simple Gaussian profile still doesn't account for the outer shell. Thus, we added a radially decreasing power law profile to the previous Gaussian profile to simulate it. All three profiles were used to test the dependence of the stellar parameters upon the choice of density profile. Figure~\ref{denshape} illustrates the different profiles.
\\\\Since each model assumes a specific value for each stellar and nebular parameter, locating the global minimum (in parameter space) or best value for each stellar parameter was helped by computing a suite of constant density models spanning a wide range of stellar and nebular parameters, using resources from the OU Supercomputing Center for Education and Research at the University of Oklahoma. To minimize the number of models required to cover the relevant parameter space, we used the following method of analysis. A primary grid of 81,000 models was produced by varying the stellar temperature and luminosity, the inner and outer radii, and the filling factor of the gas while holding the abundances of He, C, Ne, and O constant. The range of the stellar temperature and luminosity was between 50,000-100,000K ($\bigtriangleup$T = 1000K) and (log[L/L\textsubscript{\astrosun}])=3.0-5.0 ($\bigtriangleup$dex = 0.1), respectively. These ranges cover all published values of these stellar parameters. The filling factor was varied between 0.01 and 0.5, a range which encompasses typical values of planetary nebulae, in steps of 0.01. The inner/outer radius were varied from 0.039pc/0.0415pc to 0.041pc/0.0437pc with a step size of 10\textsuperscript{-3}pc/10\textsuperscript{-4}pc, covering values that are reasonable for the assumed distance. The nebular composition was chosen to be the full region's values from Table~\ref{tab:abun1}, since all regions had elemental abundances within error of the full region's values. Grains were chosen to be the \textit{planetary nebula} set internal to CLOUDY with a fixed scaling factor of 1.0.  The density was chosen to be the full region's value of 4500~cm\textsuperscript{-3} except for Regions 1, 2, 7, and 9.
\\\\Next, the set of observed emission lines strengths for NGC 3242 were used to reduce the primary grid to a smaller group of models which were the most successful at reproducing the observations, as determined by the previously discussed rms analysis. Finally, the abundances were varied within the observed errors to produce a secondary grid of over 10 million models to further reduce the model rms. Ultimately, the best values were found by starting with the best model from the grid, and manually adjusting each input parameter and  keeping the values that lowered the model rms. An estimation of the error for each parameter was carried out by starting with the best model and varying each parameter one at a time (the stellar temperature and luminosity were varied together) until the model rms was larger than the sum of the best fit model rms and observed rms\footnotemark, e.g. model rms $>$ 0.21 for Region Full. This estimation process was used only for the constant density models, due to constraints imposed by computational time requirements.
\footnotetext[3]{Admittedly, each parameter is not fully independent. However, varying the stellar parameters together takes into account more of the covariance.}
\\\\Tables~\ref{modline1}~\&~\ref{modline2} show how well the best constant density models for each region were able to match the observed emission line strengths. The first two columns are the wavelength and ion line identification, respectively. The next five columns separate the regions with the ratio of the model emission line strength to the observed emission line strength. Region Full has the most lines that are outside the errors listed in Tables~\ref{emisstab1}~\&~\ref{emisstab2}. This is probably due to the modeling assumptions becoming less valid compared to the smaller regions and the smaller uncertainties in the line strengths. Table~\ref{nmodline} has the same format as the previous two tables but shows the Gaussian and Gaussian with the power law density profile models of only Region Full. These models are able to match the observed emission lines about as well as the constant density model.
\\\\The parameters for the best fit constant density models for each region, along with their estimated errors, are shown in Tables~\ref{condenmods1}~\&~\ref{condenmods2}. The dust scaling factor was set to 0.077 for these models. The first column lists the stellar and nebular parameters. Only the parameters with associated errors were varied. Near the bottom we provide the electron temperature/density and ionization correction factors for oxygen and neon from the model. The total rms for both the model and the observation are also given. 
\\\\As can be seen, the model rms is smaller than the observed rms for all regions. The smaller the model rms, the better the model was able to match the line strengths and diagnostics for each region. For example, the smallest model rms of 1.63 from region 4 implies that the assumptions that went into the model were more representative of the true physical structure of this region. The models reproduced the observed temperatures and densities within their respective errors in all cases. The model ionization correction factors are smaller than their observed counterparts given in Tables~\ref{ionabun1}~\&~\ref{ionabun2} in all regions except region 1, with regions 7 and 8 having the largest differences.  
\\\\The asymmetry in the errors is due to the $1-\frac{model}{observed}$ term in the equation for the rms having a lower bound of 1. The errors in stellar temperature and luminosity and nebular abundances are inversely related to overall signal strength in the emission lines, where the larger error corresponds to overall weaker signal strength. Since the full region possesses the highest signal to noise ratio, we adopt its values of 89.7$^{+7.3}_{-4.7}$~kK and log(L/$L_{\astrosun}$)=3.36$^{+0.28}_{-0.22}$ for the central star of NGC 3242. Also, the model abundances in Table~\ref{condenmods1}~\&~\ref{condenmods2} agree within estimated error with the observed abundances in Table~\ref{tab:abun1}, although carbon is systematically higher and neon is systematically lower in all cases. 
\\\\Table~\ref{gaudenmods1} shows the central star parameters for the best fit models of the Gaussian and Gaussian with a power law density profiles along with the abundances and filling factor. The electron temperature and density, oxygen and neon ionization correction factors, and associated model rms are at the bottom. The first column identifies the nebular density profile, while the second column indicates the model parameters. The remaining column shows the value of each parameter for Region Full. The largest change in temperature and luminosity between the constant (Table~\ref{condenmods1}) and non-constant (Table~\ref{gaudenmods1}) density models is 1900 K and 0.1 dex, respectively, from the Gaussian density profile. The Gaussian with a power law density profile more closely matches the stellar parameters of the constant density profile and is more representative of the nebular structure. Thus, there appears to be only a minor difference between these models and the constant density models in terms of the luminosity and temperature of the central star when considering the nebula as a whole. Therefore, our adoption above for the central star parameters from the constant density models is reasonable. The filling factor, on the other hand, is much larger in both the Gaussian and Gaussian with a power law density models when compared to the constant density models. Both these profiles reproduce nearly the same electron temperature/density and ionization correction factors as the constant density profile. 

\section{Discussion}

\begin{figure}
\caption{Log \textit{L/L$_{sun}$} vs. log \textit{T$_{eff}$} for NGC 3242. Post-AGB model tracks from \citet{vas94} (solid red lines, Z=0.016), \citet{sch83} (blue dashed lines, Z=0.016) and \cite{ber16} (purple dash-dot lines, Z=0.01 and green dash lines, Z=0.02) and associated initial/final masses are shown at the right end of each track.}
\includegraphics{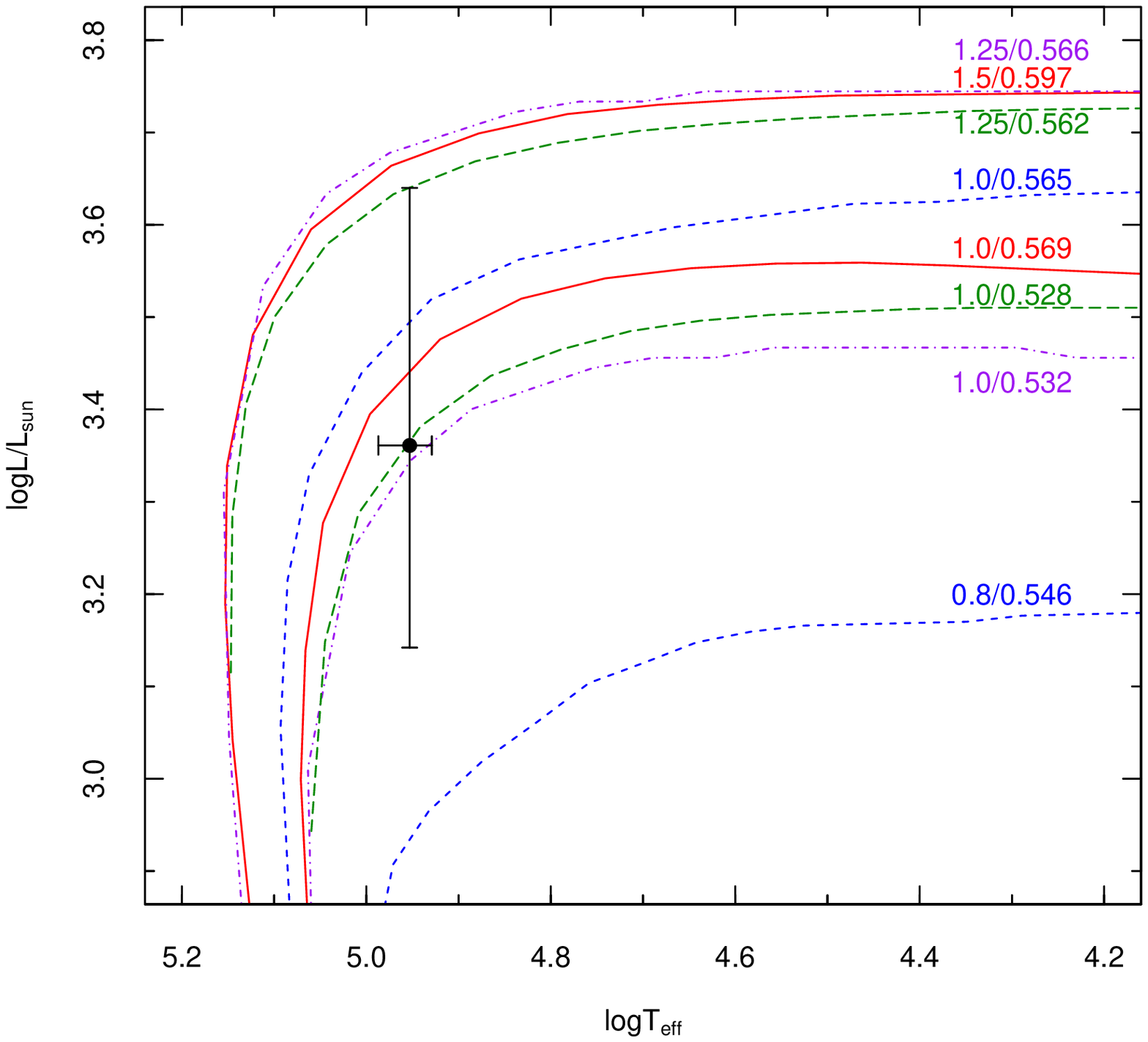}
\label{agbmods}
\end{figure}

One of our primary results is that our spatially resolved studies of various regions of NGC 3242 show no evidence of abundance variation for He, C, O, and Ne within our uncertainties (Figure~\ref{fig:abun} and Table~\ref{tab:abun1}). This result has been demonstrated here for the first time at sub-arcsecond spatial resolution. \citet{per98} performed a spatial analysis similar to ours at 1.5'' resolution on thirteen bipolar planetary nebulae and saw no inhomogeneities in the observed elements of He, O and N. However, there was an implied systematic increase in some elements (Ne, Ar and S) at larger distances from the central stars. On the other hand, \citet{bal94} found abundance variations in nitrogen for a few planetary nebulae when comparing low ionization structures to the rest of the nebula. These results were not subsequently confirmed using more sophisticated analyses \citep{gon06}.
\\\\The apparent lack of abundance variations across NGC 3242 supports the idea that the visible nebula was formed by chemically uniform outflows ejected from the central star. AGB stellar evolution models computed by \citet{ren81} indicated that the mixing timescales of convection in the envelope are much shorter than the nuclear timescales. More recent modeling by Buell (1997 and private communication) suggests that the mixing time frame is on the order of 1 yr compared to the nuclear timescale of order 10$^{5}$-10$^{6}$ yr. Our results confirm that elements such as $^{12}$C, $^{13}$C and $^{14}$N that are formed during the AGB phase of NGC 3242 are mixed into the envelope at a faster rate than mass ejections occur. 
\\\\Looking at the central star parameters, Table~\ref{tlstar} compares our best fit temperature and luminosity of the constant density models to values found in the literature of the last 25 years. Taking into account the uncertainty for both parameters in Table~\ref{condenmods1}, our temperature and luminosity are consistent with the values from~\cite{fre08} and Paper II.~\cite{tin02} has a consistent temperature but higher luminosity when compared with our results. 
\\\\From our model-derived central star temperature and luminosity, we can estimate the initial and final masses along with the current radius for NGC 3242. Figure~\ref{agbmods} is a theoretical H-R diagram with post-AGB evolutionary tracks from \citet{vas94} (red solid lines), \citet{sch83} (blue dash lines), and \cite{ber16} (green dash and purple dash-dot lines). The initial/final masses relative to the Sun can be seen at the right end of each track. The position of NGC 3242 is shown. We interpolated between evolutionary tracks and took the average of the models to obtain an initial and final mass of 0.95$^{+0.35}_{-0.09}$ M$_{\astrosun}$ and 0.56$^{+0.01}_{-0.01}$ M$_{\astrosun}$, respectively. This final mass corresponds with the peak of the distribution for white dwarf masses in the Milky Way shown by \citet{lie05}, 0.565$M_{\astrosun}$. It is also somewhat smaller than the average white dwarf mass, 0.64 M$_{\astrosun}$, found by \citet{lie05} as well as the peak central star mass, 0.60 M$_{\astrosun}$, found by \citet{zha93} in their sample. Finally, we calculated a central star radius of 0.20$^{+0.10}_{-0.07}$ R$_{\astrosun}$ using the Stefan-Boltzmann law.
\\\\We also estimated the main sequence lifetime plus ZAMS age, luminosity, radius, temperature, and spectral type of NGC 3242 from the initial mass. Using the mass-luminosity relation, L $\approx$ 1.02*M$^{3.92}$ in solar units from \citet{dem91}, the main sequence lifespan of the Sun, $\tau_{\astrosun}\approx$ M$_{\astrosun}$/L$_{\astrosun}$ =  9.5 Gyr, and the age of the Milky Way Galaxy, 13.2 Gyr, the initial luminosity and main sequence lifespan are 0.84$^{+1.98}_{-0.27}$ L$_{\astrosun}$ and 10.7$^{+2.5}_{-6.4}$ Gyr, respectively. This luminosity suggests a spectral type of G4, with a range of G7-F5 (based on the uncertainties). The radius was calculated with the mass-radius relation, R $\approx$ 1.06*M$^{0.945}$, also from \citet{dem91} and was found to be 1.01$^{+0.34}_{-0.09}$ R$_{\astrosun}$. From the Stefan-Boltzmann law, the temperature was determined to be 5500$^{+900}_{-300}$ K.

\section{Summary and Conclusions}
In this work, we began by looking for variations in the nebular properties of NGC 3242. This was accomplished by first dividing the co-spatial spectrophotometric measurements obtained during the HST Cycle 19 program GO 12600 into ten spatial regions, one full region and nine subsets of the full region. The spectra were then extracted by an in-home Python script, and line strengths were measured using the IRAF task \textit{splot}. Lastly, nebular properties were calculated by the program ELSA and used for comparing the spatial regions. 
\\\\We then aimed to constrain the stellar temperature and luminosity by modeling NGC 3242 with the program CLOUDY. We did this by using our empirically determined nebular properties along with the observed line strengths as constraints. The model-generated volume emissivities for each line were then input into the code PANIC, which converted the emissivities into line-of-sight line strengths. These lines as well as nebular diagnostics were then compared with their observed counterparts and an rms value, quantifying the closeness of the match, was used to choose the best fit model and errors in each parameter. Lastly, we tested the effects of three different density profiles on the stellar parameters by using a constant, Gaussian, and Gaussian with a power law profile for each region. 
\\\\The following conclusions emerge from our work. 
\begin{itemize}
\item The inner shell and, with lesser confidence, the outer shell of NGC 3242 are chemically homogeneous, implying that the outflow from the star that produced the shells was well-mixed.
\item The constant density models constrain the stellar temperature and, to a lesser degree, the luminosity. Changing the density profile from a constant to the Gaussian or Gaussian with a power law resulted in a large increase in the filling factor but had a negligible effect on the inferred stellar properties. The Gaussian with a power law profile was the most representative of the structure of NGC 3242 out of the three profiles tested here.
\item The progenitor mass of the central star of NGC 3242 was 0.95$^{+0.35}_{-0.09}$ M$_{\astrosun}$ when it formed $10.7^{+2.5}_{-6.4}$ Gyr ago. It had a luminosity, temperature and radius of 0.84$^{+1.98}_{-0.27}$ L$_{\astrosun}$, $5500^{+900}_{-300}$ K and 1.01$^{+0.34}_{-0.09}$ R$_{\astrosun}$, respectively. Currently, the central star has a luminosity of log(L/$L_{\astrosun}$)=3.36$^{+0.28}_{-0.22}$ with a temperature of 89.7$^{+7.3}_{-4.7}$~kK and radius 0.20$^{+0.10}_{-0.07}$ R$_{\astrosun}$.
\end{itemize}
In a follow-up paper (Paper IV: Miller et al. 2016, in prep.), we will analyze six additional objects, IC 2165, IC 3568, NGC 2440, NGC 5315, NGC 5882 and NGC 7662, as we did here in Paper III for NGC 3242.


\acknowledgments

Support for Program number GO12600 was provided by NASA through a grant from the Space Telescope Science Institute, which is operated by the Association of Universities for Research in Astronomy, Incorporated, under NASA contract NAS5-26555. All authors are grateful to their home institutions for travel support, if provided. Most of the computing for this project was performed at the OU Supercomputing Center for Education and Research (OSCER) at the University of Oklahoma. Special thanks to Jim Buell for sharing his expertise on the evolution of AGB/post-AGB stars and to Christophe Morisset whose careful review greatly improved the quality of this paper.

\clearpage



\begin{deluxetable}{clcccccccccccccccccccc}
\tabletypesize{\small}
\setlength{\tabcolsep}{0.07in}
\tablewidth{580pt}
\rotate
\tablecaption{Fluxes and Intensities I.\label{emisstab1}}
\tablehead{
\colhead{Wave} & \colhead{} & \colhead{} &
\multicolumn{2}{c}{Region Full} &
\multicolumn{2}{c}{Region 1} &
\multicolumn{2}{c}{Region 2} &
\multicolumn{2}{c}{Region 3} &
\multicolumn{2}{c}{Region 4} \\
\cline{4-13}\\
\colhead{(\AA)} &
\colhead{ID} &
\colhead{f($\lambda$)} &
\colhead{F($\lambda$)} &
\colhead{I($\lambda$)} &
\colhead{F($\lambda$)} &
\colhead{I($\lambda$)} &
\colhead{F($\lambda$)} &
\colhead{I($\lambda$)} &
\colhead{F($\lambda$)} &
\colhead{I($\lambda$)} &
\colhead{F($\lambda$)} &
\colhead{I($\lambda$)}}
\startdata
1485 & N IV] & 1.231 & 8.47 & 10.1$\pm$0.86 & 2.24 & 2.66$\pm$0.92 & 7.06 & 8.41$\pm$1.43 & 11.2 & 12.3$\pm$1.89 & 7.73 & 9.80$\pm$1.88 \\
1907 & C III] & 1.226 & 121 & 144$\pm$5 & 117 & 138$\pm$27 & 118 & 141$\pm$12 & 130 & 143$\pm$15 & 129 & 163$\pm$21 \\
1909 & C III] & 1.229 & 88.8 & 106$\pm$4 & 80.8 & 95.6$\pm$19.07 & 91.3 & 109$\pm$9 & 92.7 & 102$\pm$11 & 92.5 & 117$\pm$16 \\
3727\tablenotemark{a} & [O II] & 0.292 & 4.38 & 4.57$\pm$1.85 & \nodata & \nodata & \nodata & \nodata & \nodata & \nodata & \nodata & \nodata \\
3869 & [Ne III] & 0.252 & 96.6 & 100$\pm$2 & 106 & 110$\pm$9 & 101 & 105$\pm$7 & 90.8 & 92.7$\pm$5.26 & 105 & 110$\pm$7 \\
3969 & [Ne III] & 0.224 & 28.7 & 29.6$\pm$1.64 & 31.9 & 32.9$\pm$5.08 & 29.1 & 30.1$\pm$2.93 & 24.8 & 25.2$\pm$2.95 & 32.5 & 33.9$\pm$4.36 \\
4101 & H$\delta$ & 0.188 & 26.8 & 27.6$\pm$1.70 & 28.9 & 29.7$\pm$4.23 & 28.1 & 28.9$\pm$2.20 & 25.7 & 26.1$\pm$3.72 & 29.7 & 30.8$\pm$4.29 \\
4340 & H$\gamma$ & 0.124 & 45.5 & 46.3$\pm$1.07 & 47.3 & 48.1$\pm$4.93 & 44.7 & 45.5$\pm$1.75 & 43.1 & 43.5$\pm$2.62 & 48.0 & 49.1$\pm$3.66 \\
4363 & [O III] & 0.118 & 12.8 & 13.0$\pm$0.96 & 12.8 & 13.0$\pm$4.58 & 14.1 & 14.4$\pm$1.33 & 12.7 & 12.9$\pm$2.20 & 15.2 & 15.6$\pm$2.97 \\
4686 & He II & 0.036 & 47.4 & 47.6$\pm$0.73 & 18.6 & 18.7$\pm$3.90 & 41.1 & 41.3$\pm$2.99 & 55.8 & 56.0$\pm$2.49 & 36.6 & 36.9$\pm$3.97 \\
4861 & H$\beta$ & 0.000 & 100 & 100$\pm$0 & 100 & 100$\pm$0 & 100 & 100$\pm$0 & 100 & 100$\pm$0 & 100 & 100$\pm$0 \\
4959 & [O III] & -0.030 & 414 & 412$\pm$3 & 459 & 457$\pm$19 & 441 & 439$\pm$8 & 369 & 368$\pm$8 & 448 & 446$\pm$12 \\
5007 & [O III] & -0.042 & 1233 & 1225$\pm$8 & 1367 & 1359$\pm$53 & 1313 & 1305$\pm$22 & 1098 & 1095$\pm$22 & 1336 & 1326$\pm$34 \\
5876 & He I & -0.231 & 9.71 & 9.39$\pm$0.66 & 15.3 & 14.8$\pm$3.44 & 13.0 & 12.6$\pm$2.41 & 9.15 & 8.98$\pm$3.38 & 10.4 & 9.98$\pm$3.82 \\
6563 & H$\alpha$ & -0.360 & 297 & 282$\pm$0 & 298 & 283$\pm$2 & 297 & 282$\pm$1 & 290 & 282$\pm$1 & 302 & 282$\pm$1 \\
6678\tablenotemark{a} & He I & -0.380 & 2.28 & 2.16$\pm$0.48 & \nodata & \nodata & \nodata & \nodata & \nodata & \nodata & \nodata & \nodata \\
7136\tablenotemark{a} & [Ar III] & -0.453 & 6.83 & 6.39$\pm$0.68 & \nodata & \nodata& \nodata & \nodata & \nodata & \nodata & \nodata & \nodata \\
9532\tablenotemark{a} & [S III] & -0.632 & 10.6 & 9.67$\pm$1.91 & \nodata & \nodata & \nodata & \nodata & \nodata & \nodata & \nodata & \nodata \\
\tableline
c\tablenotemark{b} & & & 0.06 & & 0.06 & & 0.06 & & 0.03 & & 0.08 \\
H$\alpha$/H$\beta$\tablenotemark{c} & & & 2.82 & & 2.83 & & 2.82 & & 2.82 & & 2.82 \\
log F$_{H\beta}$\tablenotemark{d} & & & -12.00 & & -13.03 & & -12.79 & & -12.84 & & -13.00 \\
\enddata
\tablenotetext{a}{This line was reliably measured for Region Full only.}
\tablenotetext{b}{$\mathrm{L}$ogarithmic extinction at H$\beta$}
\tablenotetext{c}{Expected intrinsic H$\alpha$/H$\beta$ ratio at nebular temperature and density}
\tablenotetext{d}{ergs\ cm$^{-2}$ s$^{-1}$ in our extracted spectra}
\end{deluxetable}

\begin{deluxetable}{clcccccccccccccccccccc}
\tabletypesize{\small}
\setlength{\tabcolsep}{0.07in}
\tablewidth{580pt}
\rotate
\tablecaption{Fluxes and Intensities II.\label{emisstab2}}
\tablehead{
\colhead{Wave} & \colhead{} & \colhead{} &
\multicolumn{2}{c}{Region 5} &
\multicolumn{2}{c}{Region 6} &
\multicolumn{2}{c}{Region 7} &
\multicolumn{2}{c}{Region 8} &
\multicolumn{2}{c}{Region 9} \\
\cline{4-13} \\
\colhead{(\AA)} &
\colhead{ID} &
\colhead{f($\lambda$)} &
\colhead{F($\lambda$)} &
\colhead{I($\lambda$)} &
\colhead{F($\lambda$)} &
\colhead{I($\lambda$)} &
\colhead{F($\lambda$)} &
\colhead{I($\lambda$)} &
\colhead{F($\lambda$)} &
\colhead{I($\lambda$)} &
\colhead{F($\lambda$)} &
\colhead{I($\lambda$)}}
\startdata
1485 & N IV] & 1.231 & 8.22 & 10.2$\pm$2.04 & 8.85 & 11.1$\pm$2.23 & 9.31 & 11.6$\pm$1.77 & 11.6 & 13.4$\pm$1.82 & 7.47 & 9.18$\pm$2.88 \\
1907 & C III] & 1.226 & 121 & 150$\pm$18 & 123 & 154$\pm$17 & 114 & 141$\pm$18 & 122 & 141$\pm$15 & 120 & 147$\pm$20 \\
1909 & C III] & 1.229 & 88.5 & 110$\pm$14 & 90.0 & 113$\pm$13 & 91.7 & 114$\pm$14 & 88.7 & 102$\pm$11 & 82.3 & 101$\pm$14 \\
3869 & [Ne III] & 0.252 & 96.2 & 101$\pm$6 & 103 & 108$\pm$6 & 92.0 & 96.2$\pm$6.20 & 93.9 & 96.8$\pm$4.97 & 103 & 108$\pm$7 \\
3969 & [Ne III] & 0.224 & 28.6 & 29.7$\pm$3.34 & 31.1 & 32.4$\pm$3.98 & 25.4 & 26.4$\pm$3.92 & 27.5 & 28.3$\pm$2.78 & 27.8 & 28.9$\pm$5.65 \\
4101 & H$\delta$ & 0.188 & 23.8 & 24.6$\pm$3.39 & 25.5 & 26.4$\pm$3.53 & 24.3 & 25.2$\pm$3.78 & 26.1 & 26.7$\pm$2.52 & 24.1 & 24.9$\pm$3.32 \\
4340 & H$\gamma$ & 0.124 & 43.4 & 44.4$\pm$3.46 & 46.1 & 47.2$\pm$3.71 & 44.3 & 45.3$\pm$3.46 & 47.2 & 47.9$\pm$2.46 & 45.3 & 46.3$\pm$2.88 \\
4363 & [O III] & 0.118 & 12.7 & 12.9$\pm$3.23 & 13.3 & 13.6$\pm$3.61 & 12.5 & 12.8$\pm$3.17 & 14.2 & 14.4$\pm$2.19 & 14.2 & 14.5$\pm$2.08 \\
4686 & He II  & 0.036 & 47.6 & 47.9$\pm$3.26 & 47.5 & 47.9$\pm$4.24 & 52.4 & 52.7$\pm$4.33 & 68.5 & 68.8$\pm$2.44 & 47.0 & 47.3$\pm$4.78 \\
4861 & H$\beta$ & 0.000 & 100 & 100$\pm$0 & 100 & 100$\pm$0 & 100 & 100$\pm$0 & 100 & 100$\pm$0 & 100 & 100$\pm$0 \\
4959 & [O III] & -0.030 & 420 & 418$\pm$10 & 422 & 420$\pm$9 & 396 & 394$\pm$10 & 364 & 363$\pm$8 & 439 & 437$\pm$12 \\
5007 & [O III] & -0.042 & 1251 & 1242$\pm$29 & 1258 & 1248$\pm$27 & 1179 & 1171$\pm$29 & 1086 & 1080$\pm$23 & 1309 & 1300$\pm$34 \\
5876 & He I & -0.231 & 8.03 & 7.71$\pm$2.37 & 7.64 & 7.32$\pm$2.51 & 7.75 & 7.44$\pm$2.01 & 5.66 & 5.50$\pm$1.46 & 10.6 & 10.2$\pm$3.55 \\
\\
6563 & H$\alpha$ & -0.360 & 301 & 282$\pm$2 & 302 & 282$\pm$2 & 300 & 282$\pm$2 & 293 & 281$\pm$1 & 300 & 282$\pm$1 \\
\tableline
c\tablenotemark{a} & & & 0.08 & & 0.08 & & 0.08 & & 0.05 & & 0.07 \\
H$\alpha$/H$\beta$\tablenotemark{b} & & & 2.82 & & 2.82 & & 2.82 & & 2.81 & & 2.82 \\
log F$_{H\beta}$\tablenotemark{c} & & & -13.03 & & -12.99 & & -12.98 & & -12.86 & & -13.17 \\
\enddata
\tablenotetext{a}{$\mathrm{L}$ogarithmic extinction at H$\beta$}
\tablenotetext{b}{Expected intrinsic H$\alpha$/H$\beta$ ratio at nebular temperature and density}
\tablenotetext{c}{ergs\ cm$^{-2}$ s$^{-1}$ in our extracted spectra}
\end{deluxetable}



\begin{deluxetable}{cccccc}
\tabletypesize{\footnotesize}
\setlength{\tabcolsep}{0.07in}
\tablewidth{0in}
\tablecaption{Ionic Abundances\tablenotemark{a}, Temperatures and Densities\label{ionabun1}}
\tablehead{
\colhead{Ion} &
\colhead{Region Full} &
\colhead{Region 1} &
\colhead{Region 2} &
\colhead{Region 3} &
\colhead{Region 4} 
}
\startdata
He$^{+}$(5876) & 6.25$\pm$0.45(-2) & 0.106$\pm$0.026 & 8.10$\pm$1.56(-2) & 5.99$\pm$2.27(-2) & 6.61$\pm$2.55(-2)\\
He$^{+2}$(4686) & 4.40$\pm$0.07(-2) & 1.73$\pm$0.36(-2) & 3.82$\pm$0.28(-2) & 5.16$\pm$0.23(-2) & 3.40$\pm$0.37(-2)\\
icf(He) & 1.0 & 1.0 & 1.0 & 1.0 & 1.0\\
O$^{+}$(3727) & 2.23$\pm$0.90(-6) & \nodata & \nodata & \nodata & \nodata \\
O$^{+2}$(5007) & 2.58$\pm$0.21(-4) & 3.17$\pm$1.21(-4) & 2.67$\pm$0.27(-4) & 2.04$\pm$0.40(-4) & 2.48$\pm$0.53(-4)\\
O$^{+2}$(4959) & 2.59$\pm$0.21(-4) & 3.18$\pm$1.22(-4) & 2.68$\pm$0.27(-4) & 2.05$\pm$0.40(-4) & 2.48$\pm$0.54(-4)\\
O$^{+2}$(4363) & 2.58$\pm$0.21(-4) & 3.17$\pm$1.21(-4) & 2.67$\pm$0.27(-4) & 2.04$\pm$0.40(-4) & 2.48$\pm$0.53(-4)\\
O$^{+2}$(adopt) & 2.59$\pm$0.21(-4) & 3.17$\pm$1.21(-4) & 2.67$\pm$0.27(-4) & 2.04$\pm$0.40(-4) & 2.48$\pm$0.53(-4)\\
icf(O) & 1.70$\pm$0.05 & 1.16$\pm$0.05 & 1.47$\pm$0.10 & 1.86$\pm$0.33 & 1.52$\pm$0.2\\
C$^{+}$(2325)  & 3.72$\pm$0.28(-7)  & 3.20$\pm$1.09(-7)  & 3.43$\pm$0.66(-7)  & 3.50$\pm$0.70(-7)  & 4.26$\pm$1.19(-7)\\
C$^{+2}$(1909)  & 2.01$\pm$0.35(-4)  & 2.35$\pm$1.96(-4)  & 1.87$\pm$0.42(-4)  & 1.51$\pm$0.62(-4)  & 1.73$\pm$0.80(-4)\\
C$^{+2}$(1907)  & 2.01$\pm$0.35(-4)  & 2.35$\pm$1.96(-4)  & 1.87$\pm$0.42(-4)  & 1.51$\pm$0.62(-4)  & 1.73$\pm$0.80(-4)\\
C$^{+2}$(adopt)  & 2.01$\pm$0.35(-4)  & 2.35$\pm$1.96(-4)  & 1.87$\pm$0.42(-4)  & 1.51$\pm$0.62(-4)  & 1.73$\pm$0.80(-4)\\
C$^{+3}$(1549)  & 3.12$\pm$0.67(-5)  & 3.08$\pm$3.14(-5)  & 3.36$\pm$0.90(-5)  & 2.54$\pm$1.27(-5)  & 2.09$\pm$1.17(-5)\\
Ne$^{+2}$(3869)  & 5.60$\pm$0.53(-5)  & 6.89$\pm$3.06(-5)  & 5.65$\pm$0.75(-5)  & 4.50$\pm$1.02(-5)  & 5.36$\pm$1.34(-5)\\
Ne$^{+4}$(1575)  & 3.52$\pm$0.45(-5)  & 2.66$\pm$1.61(-5)  & 4.23$\pm$0.99(-5)  & 2.40$\pm$0.75(-5)  & 2.63$\pm$1.20(-5)\\
icf(Ne) &  1.72$\pm$0.05  & 1.16$\pm$0.05  & 1.47$\pm$0.10  & 1.86$\pm$0.33  & 1.52$\pm$0.20\\
\tableline
\[[O III] T$_{e}$ (K) & 11700$\pm$300 & 11400$\pm$1400 & 11900$\pm$400 & 12200$\pm$800  & 12200$\pm$800 \\
C III] N$_{e}$ ($cm^{-3}$) & 4500$\pm$300 & 1800$\pm$1700 & 7000$\pm$900 & 3300$\pm$1000 & 3700$\pm$1400 \\
\enddata
\tablenotetext{a}{\footnotesize Abundances relative to H$^{+}$; n.nn$\pm$n.nn(-k) == (n.nn$\pm$n.nn) x 10$^{-k}$}
\end{deluxetable}

\begin{deluxetable}{cccccc}
\tabletypesize{\footnotesize}
\setlength{\tabcolsep}{0.07in}
\tablewidth{0in}
\tablecaption{Ionic Abundances\tablenotemark{a}, Temperatures and Densities\label{ionabun2}}
\tablehead{
\colhead{Ion} &
\colhead{Region 5} &
\colhead{Region 6} &
\colhead{Region 7} &
\colhead{Region 8} &
\colhead{Region 9} 
}
\startdata
He$^{+}$(5876)  & 5.15$\pm$1.60(-2)  & 4.84$\pm$1.68(-2)  & 4.75$\pm$1.31(-2)  & 3.60$\pm$0.96(-2)  & 7.26$\pm$2.62(-2)\\
He$^{+2}$(4686)  & 4.43$\pm$0.30(-2)  & 4.42$\pm$0.39(-2)  & 4.87$\pm$0.40(-2)  & 6.34$\pm$0.22(-2)  & 4.36$\pm$0.44(-2)\\
icf(He) & 1.0 & 1.0 & 1.0 & 1.0 & 1.0\\
O$^{+2}$(5007)  & 2.67$\pm$0.74(-4)  & 2.54$\pm$0.76(-4)  & 2.45$\pm$0.68(-4)  & 1.96$\pm$0.35(-4)  & 2.55$\pm$0.40(-4)\\
O$^{+2}$(4959)  & 2.67$\pm$0.74(-4)  & 2.55$\pm$0.76(-4)  & 2.46$\pm$0.68(-4)  & 1.96$\pm$0.35(-4)  & 2.55$\pm$0.40(-4)\\
O$^{+2}$(4363)  & 2.67$\pm$0.74(-4)  & 2.54$\pm$0.76(-4)  & 2.45$\pm$0.68(-4)  & 1.96$\pm$0.35(-4)  & 2.55$\pm$0.40(-4)\\
O$^{+2}$(adopt)  & 2.67$\pm$0.74(-4)  & 2.55$\pm$0.76(-4)  & 2.45$\pm$0.68(-4)  & 1.96$\pm$0.35(-4)  & 2.55$\pm$0.40(-4)\\
icf(O) &  1.86$\pm$0.27 &  1.91$\pm$0.33 &  2.02$\pm$0.29 &  2.76$\pm$0.47 &  1.60$\pm$0.23\\
C$^{+}$(2325)  & 3.46$\pm$0.72(-7)  & 4.51$\pm$1.05(-7)  & 3.78$\pm$1.34(-7)  & 3.76$\pm$1.20(-7)  & 3.13$\pm$1.69(-7)\\
C$^{+2}$(1909)  & 2.18$\pm$1.30(-4)  & 1.99$\pm$1.26(-4)  & 2.01$\pm$1.19(-4)  & 1.40$\pm$0.53(-4)  & 1.69$\pm$0.59(-4)\\
C$^{+2}$(1907)  & 2.17$\pm$1.30(-4)  & 1.99$\pm$1.26(-4)  & 2.01$\pm$1.19(-4)  & 1.40$\pm$0.53(-4)  & 1.69$\pm$0.59(-4)\\
C$^{+2}$(adopt)  & 2.18$\pm$1.30(-4)  & 1.99$\pm$1.26(-4)  & 2.01$\pm$1.19(-4)  & 1.40$\pm$0.53(-4)  & 1.69$\pm$0.59(-4)\\
C$^{+3}$(1549)  & 2.88$\pm$2.10(-5)  & 2.49$\pm$1.93(-5)  & 2.06$\pm$1.49(-5)  & 1.88$\pm$0.86(-5)  & 4.57$\pm$1.90(-5)\\
Ne$^{+2}$(3869)  & 5.74$\pm$1.85(-5)  & 5.81$\pm$2.00(-5)  & 5.30$\pm$1.70(-5)  & 4.53$\pm$0.93(-5)  & 5.52$\pm$1.04(-5)\\
Ne$^{+4}$(1575)  & 3.46$\pm$1.74(-5)  & 3.60$\pm$1.59(-5)  & 4.64$\pm$1.92(-5)  & 2.83$\pm$0.86(-5)  & 2.66$\pm$1.20(-5)\\
icf(Ne) &  1.86$\pm$0.27 &  1.91$\pm$0.33 &  2.02$\pm$0.29 &  2.76$\pm$0.47 &  1.60$\pm$0.23\\
\tableline
\[[O III] T$_{e}$ (K) & 11700$\pm$1000 &  11900$\pm$1100 &  11800$\pm$1000 &  12300$\pm$700 & 12000$\pm$600 \\
C III] N$_{e}$ ($cm^{-3}$) & 4400$\pm$1300 & 4500$\pm$1200 & 9000$\pm$1200 & 4000$\pm$1300 & 1500$\pm$2000 \\
\enddata
\tablenotetext{a}{\footnotesize Abundances relative to H$^{+}$; n.nn$\pm$n.nn(-k) == (n.nn$\pm$n.nn) x 10$^{-k}$}
\end{deluxetable}

\begin{deluxetable}{lcccccc}
\setlength{\tabcolsep}{0.07in}
\tablewidth{0in}
\tablecaption{Total Elemental Abundances\label{tab:abun1}}
\tablehead{
\colhead{Parameter} &
\colhead{Full Region} &
\colhead{Region 1} &
\colhead{Region 2} &
\colhead{Region 3} &
\colhead{Region 4} &
\colhead{Solar\tablenotemark{a}} 
}
\startdata
He/H (10$^{-2}$) & 10.60$^{+0.50}_{-0.50}$ & 12.30$^{+2.60}_{-2.60}$ & 11.90$^{+1.60}_{-1.60}$ & 11.1$^{+2.3}_{-2.3}$ & 10.0$^{+2.58}_{-2.58}$ & 8.51 \\
C/H (10$^{-4}$) & 2.32$^{+0.36}_{-0.36}$ & 2.66$^{+1.98}_{-1.98}$ & 2.21$^{+0.43}_{-0.43}$ & 1.76$^{+0.63}_{-0.63}$ & 1.94$^{+0.80}_{-0.80}$ & 2.69 \\
C/O & 0.523$^{+0.091}_{-0.091}$ & 0.721$^{+0.602}_{-0.602}$ & 0.562$^{+0.127}_{-0.127}$ & 0.463$^{+0.203}_{-0.203}$ & 0.517$^{+0.248}_{-0.248}$ & 0.550 \\
O/H (10$^{-4}$) & 4.44$^{+0.36}_{-0.36}$ & 3.69$^{+1.40}_{-1.40}$ & 3.93$^{+0.45}_{-0.45}$ & 3.80$^{+0.96}_{-0.96}$ & 3.75$^{+0.92}_{-0.92}$ & 4.90 \\
Ne/H (10$^{-5}$) & 9.62$^{+0.91}_{-0.91}$ & 8.02$^{+3.55}_{-3.55}$ & 8.31$^{+1.20}_{-1.20}$ & 8.37$^{+2.31}_{-2.31}$ & 8.12$^{+2.23}_{-2.23}$ & 8.51 \\
Ne/O & 0.217$^{+0.004}_{-0.004}$ & 0.217$^{+0.016}_{-0.016}$ & 0.211$^{+0.014}_{-0.014}$ & 0.220$^{+0.012}_{-0.012}$ & 0.216$^{+0.011}_{-0.011}$ & 0.174 \\
\tableline
 & Region 5 & Region 6 & Region 7 & Region 8 & Region 9 \\
He/H (10$^{-2}$) & 9.57$^{+1.63}_{-1.63}$ & 9.26$^{+1.74}_{-1.74}$ & 9.62$^{+1.38}_{-1.38}$ & 9.93$^{+1.00}_{-1.00}$ & 11.6$^{+2.70}_{-2.70}$ & 8.51 \\
C/H (10$^{-4}$) & 2.47$^{+1.32}_{-1.32}$ & 2.24$^{+1.27}_{-1.27}$ & 2.22$^{+1.20}_{-1.20}$ & 1.59$^{+0.53}_{-0.53}$ & 2.15$^{+0.62}_{-0.62}$ & 2.69 \\
C/O & 0.498$^{+0.304}_{-0.304}$ & 0.460$^{+0.300}_{-0.300}$ & 0.447$^{+0.273}_{-0.273}$ & 0.294$^{+0.119}_{-0.119}$ & 0.527$^{+0.187}_{-0.187}$ & 0.550 \\
O/H (10$^{-4}$) & 4.96$^{+1.46}_{-1.46}$ & 4.87$^{+1.56}_{-1.56}$ & 4.97$^{+1.42}_{-1.42}$ & 5.41$^{+1.25}_{-1.25}$ & 4.08$^{+0.85}_{-0.85}$ & 4.90 \\
Ne/H (10$^{-5}$) & 10.70$^{+3.60}_{-3.60}$ & 11.1$^{+4.00}_{-4.00}$ & 10.7$^{+3.50}_{-3.50}$ & 12.5$^{+3.20}_{-3.20}$ & 8.83$^{+2.06}_{-2.06}$ & 8.51 \\
Ne/O & 0.215$^{+0.013}_{-0.013}$ & 0.228$^{+0.014}_{-0.014}$ & 0.216$^{+0.014}_{-0.014}$ & 0.231$^{+0.010}_{-0.010}$ & 0.216$^{+0.012}_{-0.012}$ & 0.174 \\
\enddata
\tablenotetext{a}{\citet{asp09}}
\end{deluxetable}
\begin{deluxetable}{clccccc}
\tabletypesize{\small}
\setlength{\tabcolsep}{0.07in}
\tablewidth{580pt}
\rotate
\tablecaption{Constant Density Model Emission Lines Compared to Observations\label{modline1}}
\tablehead{
\colhead{Wave} & \colhead{} & 
\colhead{Region Full} &
\colhead{Region 1} &
\colhead{Region 2} &
\colhead{Region 3} &
\colhead{Region 4} \\
\cline{3-7}\\
\colhead{(\AA)} &
\colhead{ID} &
\colhead{Model/Observed} &
\colhead{Model/Observed} &
\colhead{Model/Observed} &
\colhead{Model/Observed} &
\colhead{Model/Observed} 
}
\startdata
1485 & N IV] & 0.999 & 1.003 & 1.001 & 1.009 & 1.001 \\
1907 & C III] & 1.047\tablenotemark{a} & 1.013 & 1.007 & 1.014 & 1.006 \\
1909\tablenotemark{d} & C III] & 1.048\tablenotemark{a} & 1.009 & 1.007 & 1.020 & 1.003 \\
3727 & [O II] & 0.954 & \nodata & \nodata & \nodata & \nodata\\
3869 & [Ne III] & 1.038\tablenotemark{a} & 1.011 & 1.003 & 1.001 & 1.005 \\
3969\tablenotemark{c} & [Ne III] & 1.057\tablenotemark{a} & 1.018 & 1.054 & 1.110 & 0.982 \\
4101\tablenotemark{c} & H$\delta$ & 0.886\tablenotemark{a} & 0.721\tablenotemark{a} & 0.880\tablenotemark{a} & 0.968 & 0.861 \\
4340\tablenotemark{c} & H$\gamma$ & 0.939\tablenotemark{a} & 0.910 & 0.999 & 1.032 & 0.955 \\
4363\tablenotemark{d} & [O III] & 1.078\tablenotemark{a} & 0.986 & 0.969 & 0.934 & 0.952 \\
4686 & He II & 0.941\tablenotemark{a} & 1.001 & 0.904\tablenotemark{a} & 0.898\tablenotemark{a} & 0.985 \\
4861 & H$\beta$ & 0.914\tablenotemark{a} & 0.931\tablenotemark{a} & 0.961\tablenotemark{a} & 0.944\tablenotemark{a} & 0.977 \\
4959\tablenotemark{d} & [O III] & 1.068\tablenotemark{a} & 1.015 & 1.006 & 0.996 & 0.989 \\
5007 & [O III] & 1.081\tablenotemark{a} & 1.028 & 1.019\tablenotemark{a} & 1.008 & 1.001 \\
5876 & He I & 0.963 & 0.996 & 0.947 & 0.975 & 1.000 \\
6563\tablenotemark{c} & H$\alpha$ & 0.918\tablenotemark{a} & 0.943\tablenotemark{a} & 0.967\tablenotemark{a} & 0.949\tablenotemark{a} & 0.974\tablenotemark{a} \\
6678\tablenotemark{b}\tablenotemark{c} & He I & 1.114 & \nodata & \nodata & \nodata & \nodata \\
7136\tablenotemark{b} & [Ar III] & 1.003 & \nodata & \nodata & \nodata & \nodata \\
9532\tablenotemark{b} & [S III] & 1.013 & \nodata & \nodata & \nodata & \nodata 
\enddata
\tablenotetext{a}{Modeled emission line intensity outside observed error bar.}
\tablenotetext{b}{This line was reliably measured in the observations for Region Full only.}
\tablenotetext{c}{This line was excluded from the rms calculation for reasons discussed in the text.}
\tablenotetext{d}{This line was only included in a diagnostic for the rms calculation for reasons discussed in the text.}
\end{deluxetable}

\begin{deluxetable}{clccccc}
\tabletypesize{\small}
\setlength{\tabcolsep}{0.07in}
\tablewidth{580pt}
\rotate
\tablecaption{Constant Density Model Emission Lines Compared to Observations\label{modline2}}
\tablehead{
\colhead{Wave} & \colhead{} & 
\colhead{Region 5} &
\colhead{Region 6} &
\colhead{Region 7} &
\colhead{Region 8} &
\colhead{Region 9} \\
\cline{3-7}\\
\colhead{(\AA)} &
\colhead{ID} &
\colhead{Model/Observed} &
\colhead{Model/Observed} &
\colhead{Model/Observed} &
\colhead{Model/Observed} &
\colhead{Model/Observed} 
}
\startdata
1485 & N IV] & 1.007 & 1.003 & 1.003 & 1.002 & 1.004 \\
1907 & C III] & 1.023 & 1.011 & 1.022 & 1.000 & 0.991 \\
1909\tablenotemark{c} & C III] & 1.025 & 1.010 & 1.027 & 0.998 & 0.995 \\
3869 & [Ne III] & 1.011 & 1.002 & 1.007 & 0.998 & 0.998 \\
3969\tablenotemark{b} & [Ne III] & 1.036 & 1.006 & 1.106 & 1.029 & 1.124 \\
4101\tablenotemark{b} & H$\delta$ & 0.969 & 0.992 & 1.008 & 0.961 & 1.053 \\
4340\tablenotemark{b} & H$\gamma$ & 0.952 & 0.976 & 1.008 & 0.953 & 1.006 \\
4363\tablenotemark{c} & [O III] & 1.067 & 1.010 & 1.097 & 0.933 & 0.987 \\
4686 & He II & 1.001 & 0.992 & 0.985 & 0.727\tablenotemark{a} & 0.957 \\
4861 & H$\beta$ & 0.886\tablenotemark{a} & 0.954\tablenotemark{a} & 0.969 & 0.960\tablenotemark{a} & 0.979 \\
4959\tablenotemark{c} & [O III] & 1.070\tablenotemark{a} & 1.017 & 1.068\tablenotemark{a} & 0.994 & 1.020 \\
5007 & [O III] & 1.084\tablenotemark{a} & 1.030\tablenotemark{a} & 1.081\tablenotemark{a} & 1.006 & 1.032\tablenotemark{a} \\
5876 & He I & 0.992 & 0.996 & 0.981 & 0.889 & 0.989 \\\\
6563\tablenotemark{b} & H$\alpha$ & 0.889\tablenotemark{a} & 0.946\tablenotemark{a} & 0.976\tablenotemark{a} & 0.966\tablenotemark{a} & 0.987\tablenotemark{a} 
\enddata
\tablenotetext{a}{Modeled emission line intensity outside observed error bar.}
\tablenotetext{b}{This line was excluded from the rms calculation for reasons discussed in the text.}
\tablenotetext{c}{This line was only included in a diagnostic for the rms calculation for reasons discussed in the text.}
\end{deluxetable}

\begin{deluxetable}{clcc}
\tabletypesize{\footnotesize}
\setlength{\tabcolsep}{0.25in}
\tablecaption{Non-constant Density Model Emission Lines Compared to Observations For Region Full\label{nmodline}}
\tablehead{
\colhead{Wave} & \colhead{} & 
\colhead{Gaussian} &
\colhead{Gaussian With Power Law} \\
\cline{3-4}\\
\colhead{(\AA)} &
\colhead{ID} &
\colhead{Model/Observed} &
\colhead{Model/Observed}  
}
\startdata
1485 & N IV] & 0.996 & 0.996\\
1907 & C III] & 1.096\tablenotemark{a} & 1.105\tablenotemark{a}\\
1909 & C III] & 1.099\tablenotemark{a} & 1.107\tablenotemark{a}\\
3727 & [O II] & 0.948 & 0.901\\
3869 & [Ne III] & 1.045\tablenotemark{a} & 1.016\\
3969 & [Ne III] & 1.064\tablenotemark{a} & 1.034\\
4101 & H$\delta$ & 0.883\tablenotemark{a} & 0.908\tablenotemark{a}\\
4340 & H$\gamma$ & 0.940\tablenotemark{a} & 0.967\tablenotemark{a}\\
4363 & [O III] & 1.133\tablenotemark{a} & 1.176\tablenotemark{a}\\
4686 & He II & 0.904\tablenotemark{a} & 0.933\tablenotemark{a}\\
4861 & H$\beta$ & 0.918\tablenotemark{a} & 0.944\tablenotemark{a}\\
4959 & [O III] & 1.094\tablenotemark{a} & 1.120\tablenotemark{a}\\
5007 & [O III] & 1.108\tablenotemark{a} & 1.134\tablenotemark{a}\\
5876 & He I & 0.916\tablenotemark{a} & 0.951\\
6563 & H$\alpha$ & 0.923\tablenotemark{a} & 0.948\tablenotemark{a}\\
6678 & He I & 1.055 & 1.095\\
7136 & [Ar III] & 1.004 & 1.028\\
9532 & [S III] & 1.024 & 1.033
\enddata
\tablenotetext{a}{Modeled emission line intensity outside observed error bar.}
\end{deluxetable}

\begin{deluxetable}{cccccc}
\tablewidth{0in}
\tablecaption{Constant Density Models\label{condenmods1}}
\tablehead{
\colhead{Parameter} &
\colhead{Region Full} &
\colhead{Region 1} &
\colhead{Region 2} &
\colhead{Region 3} &
\colhead{Region 4}
}
\startdata
$T_{star}$ (kK) & $89.7^{+7.3}_{-4.7}$ & $89.7$ & $89.7$ & $89.7$ & $89.7$  \\
$L_{star}$ (log[L/$L_{\astrosun}$]) & $3.36^{+0.28}_{-0.22}$ & $3.36$ & $3.36$ & $3.36$ & $3.36$  \\
$H_{den}$ (log[$H_{density}$]) & 3.62$^{+0.04}_{-0.06}$ & 3.25$^{+0.06}_{-0.06}$ & 3.77$^{+0.04}_{-0.03}$ & 3.50$^{+0.06}_{-0.07}$ & 3.50$^{+0.06}_{-0.07}$ \\
Inner Radius ($10^{-2}$pc) & 1.86$^{+0.74}_{-1.86}$ & 2.04$^{+1.36}_{-2.04}$ & 3.33$^{+0.12}_{-0.16}$ & 2.52$^{+0.48}_{-0.82}$ & 1.96$^{+0.74}_{-1.96}$ \\
Outer Radius ($10^{-2}$pc) & 3.80$^{+0.30}_{-0.30}$ & 10.50$^{+2.90}_{-2.40}$ & 3.90$^{+0.13}_{-0.09}$ & 4.59$^{+0.81}_{-0.59}$ & 10.20$^{+2.60}_{-2.10}$ \\
Filling Factor ($10^{-1}$) & 4.36$^{+1.14}_{-1.06}$ & 4.39$^{+1.31}_{-1.19}$ & 4.80$^{+1.00}_{-0.90}$ & 5.93$^{+1.77}_{-1.63}$ & 1.56$^{+0.54}_{-0.46}$ \\
He/H ($10^{-2}$) & 10.45$^{+4.34}_{-3.03}$ & 11.53$^{+4.34}_{-3.38}$ & 11.02$^{+3.11}_{-2.31}$ & 10.30$^{+4.15}_{-3.39}$ & 9.57$^{+4.23}_{-3.95}$ \\
C/H ($10^{-4}$) & 5.97$^{+7.52}_{-3.73}$ & 4.85$^{+4.92}_{-3.50}$ & 3.67$^{+2.35}_{-1.89}$ & 5.01$^{+5.46}_{-3.19}$ & 4.10$^{+4.03}_{-2.32}$ \\
O/H ($10^{-4}$) & 4.70$^{+2.71}_{-1.81}$ & 4.26$^{+2.35}_{-1.30}$ & 3.76$^{+1.25}_{-0.94}$ & 3.76$^{+2.41}_{-1.57}$ & 4.21$^{+2.55}_{-1.58}$ \\
Ne/H ($10^{-5}$) & 7.40$^{+6.73}_{-6.24}$ & 7.78$^{+7.36}_{-6.37}$ & 6.40$^{+3.60}_{-3.24}$ & 6.19$^{+5.03}_{-4.72}$ & 6.90$^{+5.40}_{-5.04}$ \\
\tableline
[O III] $T_{e}$ (K) & 11800 & 11100 & 11600 & 11800 & 11800 \\
C III] $N_{e}$ ($cm^{-3}$) & 4600 & 1500 & 6800 & 3500 & 3600 \\
icf (O)  & 1.57 & 1.13 & 1.22 & 1.56 & 1.51 \\
icf (Ne)  & 1.24 & 1.00 & 1.03 & 1.18 & 1.18 \\
Model RMS ($10^{-2}$)  & 6.26 & 2.40 & 3.75 & 4.46 & 1.63 \\
Observed RMS ($10^{-2}$)  & 15.02 & 20.91 & 10.68 & 17.06 & 18.57 
\enddata
\end{deluxetable}

\begin{deluxetable}{cccccc}
\tablewidth{0in}
\tablecaption{Constant Density Models\label{condenmods2}}
\tablehead{
\colhead{Parameter} &
\colhead{Region 5} &
\colhead{Region 6} &
\colhead{Region 7} &
\colhead{Region 8} &
\colhead{Region 9}
}
\startdata
$T_{star}$ (kK) & $89.7$ & $89.7$ & $89.7$ & $89.7$ & $89.7$  \\
$L_{star}$ (log[L/$L_{\astrosun}$]) & $3.36$ & $3.36$ & $3.36$ & $3.36$ & $3.36$  \\
$H_{den}$ (log[$H_{density}$]) & 3.60$^{+0.05}_{-0.06}$ & 3.60$^{+0.05}_{-0.06}$ & 3.90$^{+0.04}_{-0.04}$ & 3.55$^{+0.05}_{-0.06}$ & 3.25$^{+0.06}_{-0.07}$ \\
Inner Radius ($10^{-2}$pc) & 2.52$^{+0.48}_{-0.32}$ & 1.88$^{+0.52}_{-1.88}$ & 2.02$^{+0.08}_{-0.12}$ & 2.98$^{+0.22}_{-0.28}$ & 3.56$^{+0.84}_{-3.56}$ \\
Outer Radius ($10^{-2}$pc) & 4.21$^{+0.69}_{-0.41}$ & 6.49$^{+1.31}_{-0.96}$ & 2.52$^{+0.08}_{-0.12}$ & 3.72$^{+0.28}_{-0.22}$ & 8.40$^{+1.90}_{-1.50}$ \\
Filling Factor ($10^{-1}$) & 3.77$^{+1.03}_{-0.97}$ & 1.52$^{+0.48}_{-0.42}$ & 4.53$^{+0.87}_{-0.83}$ & 1.00$^{+0.00}_{-0.27}$ & 4.34$^{+1.36}_{-1.34}$ \\
He/H ($10^{-2}$) & 10.12$^{+3.69}_{-2.87}$ & 9.20$^{+3.68}_{-3.32}$ & 9.29$^{+2.19}_{-1.88}$ & 7.69$^{+2.78}_{-2.56}$ & 10.50$^{+4.64}_{-4.04}$ \\
C/H ($10^{-4}$) & 6.30$^{+6.29}_{-2.39}$ & 5.60$^{+5.12}_{-3.09}$ & 4.47$^{+4.04}_{-2.65}$ & 3.51$^{+3.91}_{-2.36}$ & 4.81$^{+5.19}_{-3.15}$ \\
O/H ($10^{-4}$) & 5.30$^{+2.65}_{-1.75}$ & 5.08$^{+2.68}_{-1.84}$ & 3.77$^{+1.60}_{-1.14}$ & 3.30$^{+1.94}_{-1.35}$ & 4.54$^{+2.38}_{-1.59}$ \\
Ne/H ($10^{-5}$) & 7.93$^{+6.20}_{-5.74}$ & 8.07$^{+6.38}_{-5.67}$ & 5.85$^{+3.92}_{-3.71}$ & 5.48$^{+4.52}_{-4.10}$ & 7.38$^{+6.42}_{-5.72}$ \\
\tableline
[O III] $T_{e}$ (K)  & 11500 & 11700 & 11700 & 11900 & 11500 \\
C III] $N_{e}$ ($cm^{-3}$) & 4500 & 4400 & 9200 & 3900 & 1600 \\
icf (O)  & 1.54 & 1.74 & 1.35 & 1.46 & 1.46 \\
icf (Ne)  & 1.19 & 1.30 & 1.09 & 1.07 & 1.16 \\
Model RMS ($10^{-2}$)  & 4.03 & 1.67 & 2.63 & 9.90 & 3.22 \\
Observed RMS ($10^{-2}$)  & 16.46 & 17.63 & 15.16 & 13.12 & 18.50 
\enddata
\end{deluxetable}

\begin{deluxetable}{ccc}
\tablewidth{0in}
\tablecaption{Non-constant Density Models For Region Full\label{gaudenmods1}}
\tablehead{
\colhead{Parameter} &
\colhead{Gaussian} &
\colhead{Gaussian With Power Law} 
}
\startdata
$T_{star}$ (kK) & 91600 & 91200\\
$L_{star}$ (log[L/$L_{\astrosun}$]) & 3.26 & 3.29\\
Filling Factor ($10^{-1}$) & 9.77 & 9.83\\
He/H ($10^{-2}$) & 9.84 & 9.93\\
C/H ($10^{-4}$) & 5.28 & 5.21\\
O/H ($10^{-4}$) & 4.54 & 4.46\\
Ne/H ($10^{-5}$) & 7.11 & 6.61\\
\tableline
\[O III] $T_{e}$ (K) & 11900 & 12000\\
C III] $N_{e}$ ($cm^{-3}$) & 4700 & 4700\\
icf (O) & 1.52 & 1.53\\
icf (Ne) & 1.22 & 1.23\\
RMS ($10^{-2}$) & 7.88 & 8.61
\enddata
\end{deluxetable}

\begin{deluxetable}{lcc}
\tablecaption{Central Star Parameters Comparison \label{tlstar}}
\tablewidth{0pt}
\tablehead{
\colhead{Reference} & \colhead{Temperature (kK)} & \colhead{Luminosity (log[L/L\textsubscript{\astrosun}])} 
}
\startdata
This Work & 89.7 & 3.36\\
Paper II & 89.0 & 3.64\\
\cite{fre08} & 89.0 & 3.54\\
\cite{pau04} & 75.0 & 3.51\\
\cite{tin02} & 94.5 & 3.75\\
\cite{hen00} & 60.0 & 4.30\\
\cite{kud97} & 75.0 & 4.01\\
\cite{ack92} & 60.0 & 3.48
\enddata
\end{deluxetable}




\clearpage


\begin{thebibliography}{}
\bibitem[Acker et al.(1992)]{ack92} Acker, A., Marcout, J., 
Ochsenbein, F., et al.\ 1992, The Strasbourg-ESO Catalogue of Galactic 
Planetary Nebulae.~Parts I, II., by Acker, A.; Marcout, J.; Ochsenbein, F.; 
Stenholm, B.; Tylenda, R.; Schohn, C..~ European Southern Observatory, 
Garching (Germany), 1992, 1047 p., ISBN 3-923524-41-2
\bibitem[Asplund et al.(2009)]{asp09} Asplund, M., Grevesse, N., Sauval, A.~J. \& Scott, P.\ 2009, \ The Chemical Composition of the Sun, Annual Review of Astronomy and Astrophysics, 47(1):481-522
\bibitem[Balick et al.(1994)]{bal94} Balick, B., Perinotto, 
M., Maccioni, A., Terzian, Y., \& Hajian, A.\ 1994, \apj, 424, 800 
\bibitem[Bertolami (2016)]{ber16} Bertolami, M., M., M. 2016, http://arxiv.org/abs/1512.04129v2
\bibitem[Bostroem 
\& Proffitt(2011)]{bos11} Bostroem, K.~A., \& Proffitt, C.\ 2011, STIS Data Handbook, HST Data Handbooks
\bibitem[Buell(1997)]{bue97} Buell, J.~F.\ 1997, 
Ph.D.~Thesis
\bibitem[Demircan 
\& Kahraman(1991)]{dem91} Demircan, O., \& Kahraman, G.\ 1991, \apss, 181, 313 
\bibitem[Dufour et al.(2015)]{duf15} Dufour, R.~J., Kwitter, 
K.~B., Shaw, R.~A., et al.\ 2015, \apj, 813, 121 (Paper I)
\bibitem[Ferland et al.(2013)]{fer13} Ferland, G.~J., Porter, 
R.~L., van Hoof, P.~A.~M., et al.\ 2013, \rmxaa, 49, 137
\bibitem[Frew(2008)]{fre08} Frew, D.~J.\ 2008, Ph.D.~Thesis
\bibitem[Frew et al.(2016)]{fre16} Frew, D.~J., Parker, 
Q.~A., \& Boji{\v c}i{\'c}, I.~S.\ 2016, \mnras, 455, 1459 
\bibitem[Gon{\c c}alves et al.(2006)]{gon06} Gon{\c c}alves, 
D.~R., Ercolano, B., Carnero, A., Mampaso, A., 
\& Corradi, R.~L.~M.\ 2006, \mnras, 365, 1039 
\bibitem[Henry et al.(2000)]{hen00} Henry, R.~B.~C., Kwitter, 
K.~B., \& Bates, J.~A.\ 2000, \apj, 531, 928
\bibitem[Henry et al.(2015)]{hen15} Henry, R.~B.~C., Balick, 
B., Dufour, R.~J., et al.\ 2015, \apj, 813, 121 (Paper II)
\bibitem[Johnson et al.(2006)]{joh06} Johnson, M.~D., Levitt, 
J.~S., Henry, R.~B.~C., 
\& Kwitter, K.~B.\ 2006, Planetary Nebulae in our Galaxy and Beyond, 234, 439
\bibitem[Krabbe 
\& Copetti(2006)]{kra06} Krabbe, A.~C., \& Copetti, M.~V.~F.\ 2006, \aap, 450, 159  
\bibitem[Kudritzki et al.(1997)]{kud97} Kudritzki, R.~P., 
Mendez, R.~H., Puls, J., 
\& McCarthy, J.~K.\ 1997, Planetary Nebulae, 180, 64
\bibitem[Kwitter 
\& Henry(2001)]{kwi01} Kwitter, K.~B., \& Henry, R.~B.~C.\ 2001, \apj, 562, 804 
\bibitem[Liebert et al.(2005)]{lie05} Liebert, J., Bergeron, 
P., \& Holberg, J.~B.\ 2005, \apjs, 156, 47 
\bibitem[Milingo et al.(2002)]{mil02} Milingo, J.~B., Henry, 
R.~B.~C., \& Kwitter, K.~B.\ 2002, \apjs, 138, 285 
\bibitem[Monteiro et al.(2013)]{mon13} Monteiro, H., Gon{\c c}alves, D.~R., Leal-Ferreira, M.~L., 
\& Corradi, R.~L.~M.\ 2013, \aap, 560, A102 
\bibitem[Morisset \& Georgiev(2009)]{mor09} Morisset, C., \& Georgiev, L.\ 2009, \aap, 507, 1517 
\bibitem[Pauldrach et al.(2004)]{pau04} Pauldrach, A.~W.~A., Hoffmann, T.~L., 
\& M{\'e}ndez, R.~H.\ 2004, \aap, 419, 1111 
\bibitem[Perinotto 
\& Corradi(1998)]{per98} Perinotto, M., \& Corradi, R.~L.~M.\ 1998, \aap, 332, 721
\bibitem[Rauch(2003)]{rau03} Rauch, T.\ 2003, \aap, 403, 709
\bibitem[Renzini 
\& Voli(1981)]{ren81} Renzini, A., \& Voli, M.\ 1981, \aap, 94, 175 
\bibitem[Ruiz et al.(2011)]{rui11} Ruiz, N., Guerrero, M.~A., 
Chu, Y.-H., \& Gruendl, R.~A.\ 2011, \aj, 142, 91
\bibitem[Savage 
\& Mathis(1979)]{sav79} Savage, B.~D., \& Mathis, J.~S.\ 1979, \araa, 17, 73
\bibitem[Schoenberner(1983)]{sch83} Schoenberner, D.\ 1983, 
\apj, 272, 708 
\bibitem[Seaton(1979)]{sea79} Seaton, M.~J.\ 1979, \mnras, 
187, 73P
\bibitem[Tinkler 
\& Lamers(2002)]{tin02} Tinkler, C.~M., \& Lamers, H.~J.~G.~L.~M.\ 2002, \aap, 384, 987 
\bibitem[Tsamis et al.(2003)]{tsa03} Tsamis, Y.~G., Barlow, 
M.~J., Liu, X.-W., Danziger, I.~J., 
\& Storey, P.~J.\ 2003, \mnras, 345, 186
\bibitem[Vassiliadis 
\& Wood(1994)]{vas94} Vassiliadis, E., \& Wood, P.~R.\ 1994, \apjs, 92, 125  
\bibitem[Zhang 
\& Kwok(1993)]{zha93} Zhang, C.~Y., \& Kwok, S.\ 1993, \apjs, 88, 137 
\end{thebibliography}
\end{document}